\documentstyle[11pt,epsf]{article}
\input epsf
\begin{document}
\baselineskip 100pt
\renewcommand{\baselinestretch}{1.5}
\renewcommand{\arraystretch}{0.666666666}
{\large
\parskip.2in
\newcommand{\be}{\begin{equation}}
\newcommand{\ee}{\end{equation}}
\newcommand{\br}{\bar}
\newcommand{\fr}{\frac}
\newcommand{\lm}{\lambda}
\newcommand{\ra}{\rightarrow}
\newcommand{\al}{\alpha}
\newcommand{\bt}{\beta}
\newcommand{\pr}{\partial}
\newcommand{\hs}{\hspace{5mm}}
\newcommand{\up}{\upsilon}
\newcommand{\dg}{\dagger}
\newcommand{\ve}{\varepsilon}
\newcommand{\acc}{\\[3mm]}

\hfill DTP\,98/13

\bigskip
\begin{center}
{\bf A Note on Ward's Chiral Model.}\footnote{To Appear in {\it Physics
Letter A}}
\end{center}

\bigskip
\begin{center}
T. I{\small OANNIDOU} and W.J. Z{\small AKRZEWSKI}\\
{\sl Department of Mathematical Sciences, University of Durham,\\
Durham DH1 3LE, UK}
\end{center}

{\bf Abstract.}
A one parameter generalization  of Ward's chiral model in 2+1 dimensions
is given.
 Like the original model the present one is integrable and possesses a
positive-definite and conserved energy and  $y$-momentum. 
The details of the scattering depend on the value of the parameter of the
generalisation.

In this note  we take a fresh look at Ward's integrable chiral model
in (2+1) dimensions \cite{W}, which is defined by its time evolution
equation
\be
(\eta^{\mu \nu}+\ve^{\mu \nu \al}V_\al) \pr_\mu(\pr_\nu \Psi
\,\Psi^\dg)=0.
\label{ch}
\ee
$\Psi$ is a map from ${\bf R}^{2+1}$ to $SU(2)$ and can be
thought of as a $2 \times 2$ unitary matrix valued function of coordinates 
$x^\mu=(t,x,y)$ and $^\dagger$ denotes the hermitian conjugation.
Greek indices range over the values $0, 1, 2$, $\pr_\mu$ denotes partial
differentiation with respect to $x^\mu$,
$\ve^{\mu \nu \al}$ is the totally skew tensor with $\ve^{012}=1$, and
$V_\al$ is a constant unit vector, {\it ie} $V_\al \,V^\al=1$.
Indices are raised and lowered using the Minkowski metric $\eta^{\mu
\nu}=\mbox{diag}(-1,1,1)$.

This model was derived by Ward, by dimensional reduction, from the
self-dual Yang-Mills equation in (2+2) dimensions. The dimensional 
reduction first gave him a Yang-Mills Higgs system in (2+1)
dimensions governed by the equation:
 \be
D_\mu \Phi = \fr{1}{2} \varepsilon_{\mu \al \bt} F^{\al \bt},
\label{Higg}
\ee
where $\Phi$, the Higgs field, is a function on ${\bf R}^{2+1}$ with
values in the Lie algebra $su(2)$. Its
covariant derivative is $D_\mu \Phi=\pr_\mu \Phi +[A_\mu, \Phi]$, the
SU(2) gauge potential is $A_\mu$ and the corresponding gauge field is
$F_{\mu \nu}=\pr_\mu A_\nu-\pr_\nu A_\mu+[A_\mu, A_\nu]$.
He then rewrote (\ref{Higg}) as (\ref{ch}) and having made a choice of
$V_\al$ as $V_\al=(0,1,0)$ discussed many properties of his model. 
Ward's model resembles the (2+1) dimensional reduction of the self-dual
equation introduced by Manakov and Zakharov \cite{MZ}.
In their case $V_\al=(i,0,0)$, and the behaviour of the solutions of both
models is similar.
However, in the Manakov-Zakharov model, an energy functional seems not to
exist. 
Therefore, in this paper we restrict our attention to Ward's model and
its generalization.

The choice of the  vector $V_\al$ is very important.
First of all, note that the standard $SU(2)$ chiral equation in (2+1)
dimensions has $V_\al=(0,0,0)$ and, although is Lorentz covariant, it
does not seem to be integrable.
By contrast, the existence of the vector $V_\al$ in (\ref{ch}) breaks
explicitly the Lorentz covariance of the model by picking out a particular
direction in space-time.
Actually, Ward's choice of $V_\al$ was motivated by the energy 
conservation. In \cite{W} he took
\be
T^{\mu\nu}=(-\eta^{\mu\alpha}\eta^{\nu\beta}+
\frac{1}{2}\eta^{\mu\nu}\eta^{\alpha\beta})\,\mbox{tr}(J^{-1}J_{\alpha}J^{-1}
J_{\beta}),
\label{arx}
\ee
and then showed that its divergence due to (\ref{ch}) is,
\be
\partial_{\mu}T^{\mu\nu}=-\frac{1}{3}\,V^{\nu}\,\varepsilon^
{\alpha\beta\gamma}\,\mbox{tr}(J^{-1}
J_{\alpha}J^{-1}J_{\beta}J^{-1}J_{\gamma}).
\ee
Here $\Psi(\lm=0,t,x,y)=J^{-1}(t,x,y)$ and $J_\mu=\pr_\mu J$.
So $T^{\mu 0}$ is conserved, if and only if, $V_0=0$: which is true in
this case.
Note that Ward's model has the same energy functional as the standard $SU(2)$ 
chiral equation, since the additional term in (\ref{ch}) is analogous to a
background magnetic field, and so does not affect the energy.

Equation (\ref{ch}) with Ward's choice for the unit vector  $V_\al$ has
many properties of an integrable system.  
It arises as a consistency condition for a pair of
linear equations, it passes the Painlev\'e test
\cite{Wa} for integrability and has an  inverse scattering transformation
\cite{Vil}. 
It admits multisoliton solutions \cite{W,W1,Ion,Zak-Ion} (more properly,
{\it lumps}, since they are algebraically decaying) and possesses
an infinite set of conserved quantities \cite{W-Ion}.
In this note we return to (\ref{ch}) and look at other choices
of the vector $V_\al$. In fact, we will show that most observations of Ward
can be extended to the case of 
 $V_\al=(\lm, 1, \lm)$ where $\lm$ is arbitrary. 
In this case,  $\Psi$ in (\ref{ch}) becomes a function of $\lm$; while for 
$\lm=0$ (\ref{ch}) reduces to  Ward's model.

This method is similar in nature to the dressing method introduced by
Zakharov and Shabat \cite{ZS} which has been a powerful tool for obtaining
new integrable nonlinear equations as well as characterizing large classes
of solutions of these equations.
This method is applicable to both equations in (1+1) dimensions (cf.
\cite{ZS}-\cite{ZS1}), as well as to equations in (2+1) dimensions (cf.
\cite{ZS1}-\cite{ZM1}).

System (\ref{ch}) is integrable in the sense that can be written as
the compatibility condition for the following linear system:
\begin{eqnarray}
(\zeta \pr_x -\pr_t-\pr_y) {\cal Z}=-A {\cal Z}, \nonumber \\
(\zeta \pr_t-\zeta \pr_y-\pr_x ){\cal Z}=-B {\cal Z}.
\label{lax}
\end{eqnarray}
Here $\zeta$ is a complex parameter, $A$ and $B$ are $2 \times 2$
anti-Hermitian trace-free matrices, depending only on $(t,x,y)$ but not on
$\zeta$ , and ${\cal Z}(\zeta, t,x,y)$ is a $2 \times 2$  matrix
satisfying $\mbox{det} \,{\cal Z}=1$, and the reality condition
\be
{\cal Z}(\br{\zeta}, t,x,y)^\dagger ={\cal Z}(\zeta, t,x,y)^{-1}.
\label{rea}
\ee
The system (\ref{lax}) is overdetermined  and in order for a solution
${\cal Z}$ to exist, $A$ and $B$ have to satisfy the
integrability conditions,
\be
\pr_x B=\pr_t A-\pr_y A, \hs \hs \hs \pr_x A-\pr_t B-\pr_y B+[A,B]=0.
\label{con}
\ee
If we put $\Psi(\lm,t,x,y)={\cal Z}(\zeta=\lm, t,x,y)$, we find by
comparing (\ref{lax}) and (\ref{con}) that
\begin{eqnarray}
A=-\lm \Psi_x \Psi^\dg + \Psi_t \Psi^\dg+ \Psi_y
\Psi^\dg, \nonumber\\
B=-\lm \Psi_t \Psi^\dg+ \lm \Psi_y \Psi^\dg +\Psi_x \Psi^\dg.
\label{kai}
\end{eqnarray}
Therefore, the integrability condition for (\ref{lax}) implies that there
exist a field $\Psi$ that satisfies the equation of motion (\ref{ch}); and
is unitary (due to the reality condition).

Let us return now to the Yang-Mills Higgs system (\ref{Higg}) from which
Ward's model was derived. 
Note that for this system  there exist a gauge in which the fields have
the term
\be
A_t\equiv A_y=\fr{1}{2}A, \hs \hs \hs A_x \equiv -\Phi =\fr{1}{2} B.
\label{coef}
\ee
where $A$ and $B$ given by (\ref{kai}).
Then, it is easily checked that (\ref{Higg}) is equivalent to (\ref{ch}).
This relation does not depend on the value of $\lambda$ and so it
suggests that our discussion of a more general $V_\al$ is  not misguided.

One advantage of the $\Psi$-description is that, as shown by Ward
(when $\lambda=0$), the system has a conserved, positive-definite energy
given by $T\sp{00}$ (\ref{arx}).
However, Ward's choice  holds only for $\lambda=0$.
When $\lambda\ne0$ we have to consider a different expression.
To do this we consider  
\be
 \Lambda_{\hspace{2mm} \nu}^\rho=\left( \begin{array}{llcl}
 (1+\lm^2)^{1/2} & \hspace{2.7mm} \lm(1+\lm^2)^{-1/2} &
 \hspace{2.7mm}\lm^2(1+\lm^2)^{-1/2}\acc
\hs \lm & \hs \hs \hspace{.8mm}  1 &  \lm\acc
\hs  0 & -\lm(1+\lm^2)^{-1/2} & \hspace{7mm} (1+\lm^2)^{-1/2}\\
\end{array} \right).
\ee \acc
This tensor represents a Lorentz transformation which takes
$V_\al=(\lambda,1,\lambda)$ to $V_\al=(0,1,0)$.

Then we can  define a new  energy-momentum tensor by
\be
 \Theta^{\rho \mu}= \Lambda_{\hspace{2mm} \nu}^\rho\, \, 
(-\eta^{\mu\alpha}\eta^{\nu\beta}+
\frac{1}{2}\eta^{\mu\nu}\eta^{\alpha\beta})
\,\mbox{tr}( 
\Psi_{\alpha} \Psi^\dg \Psi_\bt \Psi^\dg),   
\ee
which, if we set $\Psi(\lm)=J^{-1}$, is just a 
Lorentz transform of (\ref{arx}).
Clearly $\Theta^{\rho \mu}$ is non-symmetric in $\mu$ and $\rho$.
If we impose (\ref{ch}), the divergence of $\Theta^{\rho \mu}$ is
\be
\pr_\mu \, \Theta^{\rho \mu}=\fr{1}{3} R^\rho \,  \varepsilon^ 
{\alpha\beta\gamma}\,\mbox{tr}(
\Psi_{\alpha} \Psi^\dg \Psi_{\beta} \Psi^{\dg} \Psi_{\gamma}\Psi^{\dg}
),
\ee
with $R^\rho\equiv \Lambda^\rho_{\, \, \nu} V^\nu=(0,1,0)$.
So the conservation of $\Theta^{\rho\mu}$ mirrors the conservation of
$T\sp{\mu\nu}$ in the Ward case, and so does the energy-momentum vector
$P^\mu=\Theta^{\mu 0}$.

Consequently, we have $\pr_\mu \, \Theta^{i \mu}=0$, for $i=0,\,2$ 
and so the energy and $y$-momentum, which are the integrals of the
densities
\begin{eqnarray}
P^0&=&(\fr{1}{2}+\fr{\lm^2}{2}) \,\mbox{tr}(\Psi_t 
\Psi^\dg_t+\Psi_x \Psi^\dg_x+\Psi_y 
\Psi^\dg_y)-\lm \, \mbox{tr}(\Psi_x \Psi^\dg_t)-\lm^2 \,\mbox{tr}(\Psi_y 
\Psi^\dg_t), \nonumber\\
P^2&=&\lm \, \mbox{tr}(\Psi_x \Psi^\dg_t)-\mbox{tr}(\Psi_y
\Psi^\dg_t),
\end{eqnarray}
are well-defined  and independent of $t$.
By contrast, the $x$-momentum density is not conserved, since $R^1=1$.
In order to ensure the finiteness of the energy, we require $\Psi$ to be 
smooth and be given by 
\be
\Psi=\Psi_0+\Psi_1(\theta) \,O(r^{-1})\,+\,O(r^{-2}),
\ee
at spatial infinity; where $\Psi_0$ denotes a constant 
$SU(2)$ matrix, $\Psi_1$ is independent of $t$ and of $x+iy=r\, e^{i
\theta}$.

To obtain explicit solutions of (\ref{ch}) Ward \cite{W} used $\Psi$ of
the Lax pair formulation of his model,  and then using the standard
methods of {\it Riemann problem with zeros} derived  families of
solutions which correspond to localized lumps of energy.
This $\Psi(\lm)$ is a solution of (\ref{ch}) describing
extended structures which interact with each other trivially or
nontrivially and  we shall study the effects of the parameter $\lm$ on
this interaction.

Let us take, as an example, a solution of (\ref{ch}) representing two 
lumps which undergo a $90^0$ scattering.
In this case the solution $\Psi(\lm)$ takes the form of a product
\be
\Psi=\left(\fr{\lm-i}{\lm+i}\right) \left(1+\fr{2 \,i}{\lm-i}\,\fr{q^\dg
\otimes q}{|q|^2}\right) \left(1+\fr{2 \,i}{\lm-i}\,\fr{p^\dg \otimes
p}{|p|^2}\right),
\ee
where $q$ and $p$ are two-dimensional vectors
\begin{eqnarray}
q&=&(1+|f|^2)(1,f)-2i(tf^\prime+h)(\bar{f},-1),\nonumber \\
p&=&(1,f).
\end{eqnarray}
So we have a family of solutions depending on the value of the
real parameter $\lm$ and on two arbitrary meromorphic rational functions
$f$ and $h$ of $z=x+ i y$. 
Let us look in more detail  at the  case of $f(z)=z$ and 
$h(z)=z^2$.

Using the same arguments as in \cite{W1} we see that $\Psi$ departs
from its asymptotic value when $(tf'+h)=0$ and at this time we can
approximately identify two separate lumps.
For $t$ negative, they are on the $x$-axis at 
$x=\pm \sqrt{-t}$, while for $t$ positive, they are on the $y$-axis at 
$y=\pm \sqrt{t}$. 
So the evolution can be described as being given by  two lumps
accelerating towards each other, scattering at right angles, and then
decelerating as they separate.
In fact, their acceleration is such as if the force of their attraction
were proportional to the inverse cube of the distance between them.
Similar behaviour has recently also, been observed in some other
integrable (2+1) dimensional models (cf. \cite{AM}).
This suggests that our results are more generic in their nature.

The evolution of our lumps can be verified by looking in more
detail at their energy density, which is,
\begin{eqnarray}
P^0 &=&16 \,
\fr{1+10r^2+5r^4+4t^2(1+2r^2)-8t(x^2-y^2)}
{[1+2r^2+5r^4+4t^2+8t(x^2-y^2)]^2}\nonumber \acc
&  + &\, 64 \,\lm \,
\fr{ (1+r^2) [x(1-t)-\lm y(t+1)]} 
{(1+\lm^2)[1+2r^2+5r^4+4t^2+8t(x^2-y^2)]^2}.
 \end{eqnarray}
We see that for large (positive) $t$,  ${P\sp0}$ has maxima at two 
points on the $y$-axis, namely $y \approx \pm\sqrt{t}$.
Moreover, the height of the corresponding lumps is proportional to 
$1/t$, which means that the $y$-axis asymmetry vanishes at $t \rightarrow 
\infty$.
Plots of the energy density are given in Figure 1 for some values of $\lm$.

Obviously, the parameter $\lm$ not only deforms the size of the corresponding 
lumps but also changes their velocity.
This follows from the $y$-momentum density (see Figure 2) which is given
by
\be
P^2=64\,\fr{(1+r^2)(\lm\,x(1-t)+y(1+t))}
{(1+\lm^2)[1+2r^2+5r^4+4t^2+8t(x^2-y^2)]^2}.
\ee
Here we note the antisymmetry of $P^2$ under the interchange $x \rightarrow
-x$, $y \rightarrow -y$.

Looking at Figure 1 we note that for $\lm=2$ the lumps are deformed
in the $y$-direction. This is not very visible for $t<0$, becomes 
quite clear at $t=0$ and results, in a situation in which the lump
moving in the positive $y$-direction has larger total energy than the 
other one.
This is consistent with the behaviour of the $y$-momentum distribution.
This behaviour is generic in nature and, in general, will hold for other
lump field configurations.

In conclusion - we have presented a one parameter generalisation
of Ward's chiral model. The models possess a conserved total energy and a
conserved $y$-component of the total momentum. 
For all values of the parameter the models
are integrable and have solutions with ``solitonic-like'' properties. 
The scattering of these lumps is similar in nature but differs
in detail; {\it} depending on the value of this parameter the lumps
can be deformed at all stages of their evolution.

\begin{figure}[b]
\unitlength1cm
\begin{picture}(9,5.5)
\put(,6){(a)}
\put(2.5,4.5){$t=-1$}
\epsfxsize=12cm
\epsffile{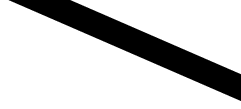 }
\end{picture}
\par
\hfill  
\begin{picture}(9.25,4.5)
\put(2.5,4.5){$t=0$} 
\epsfxsize=12cm
\epsffile{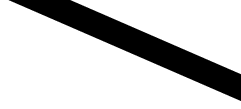}
\end{picture}
\par
\hfill
\begin{picture}(17,1.5)
\put(2.5,4.5){$t=1$}
\epsfxsize=12cm
\epsffile{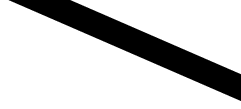}
\end{picture}
\par
\hfill
\put(,6.5){(b)}
\begin{picture}(16,5.5)
\put(2.5,4.5){$t=-2$}
\epsfxsize=12cm
\epsffile{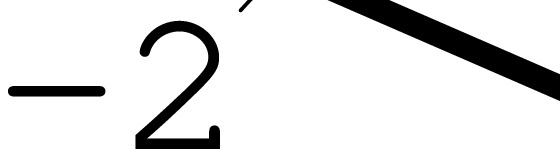}
\end{picture}
\par
\hfill
\begin{picture}(9.25,4.5)
\put(2.5,4.5){$t=0$}
\epsfxsize=12cm
\epsffile{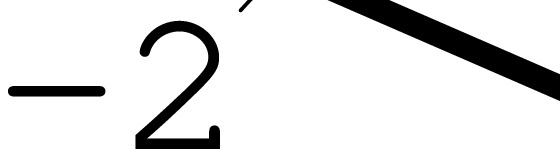}
\end{picture}
\par
\hfill
\begin{picture}(17,1.5) 
\put(2.5,4.5){$t=2$}
\epsfxsize=12cm
\epsffile{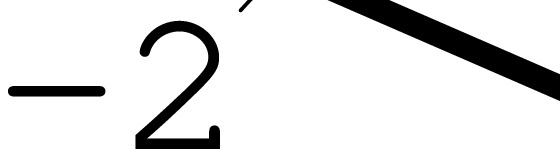}
\end{picture}     
\end{figure}

\begin{figure}[b]
\unitlength1cm
\begin{picture}(10,4.75)
\put(,5.5){(a)}
\put(2,2.75){$t=-3$}
\epsfxsize=10cm
\epsffile{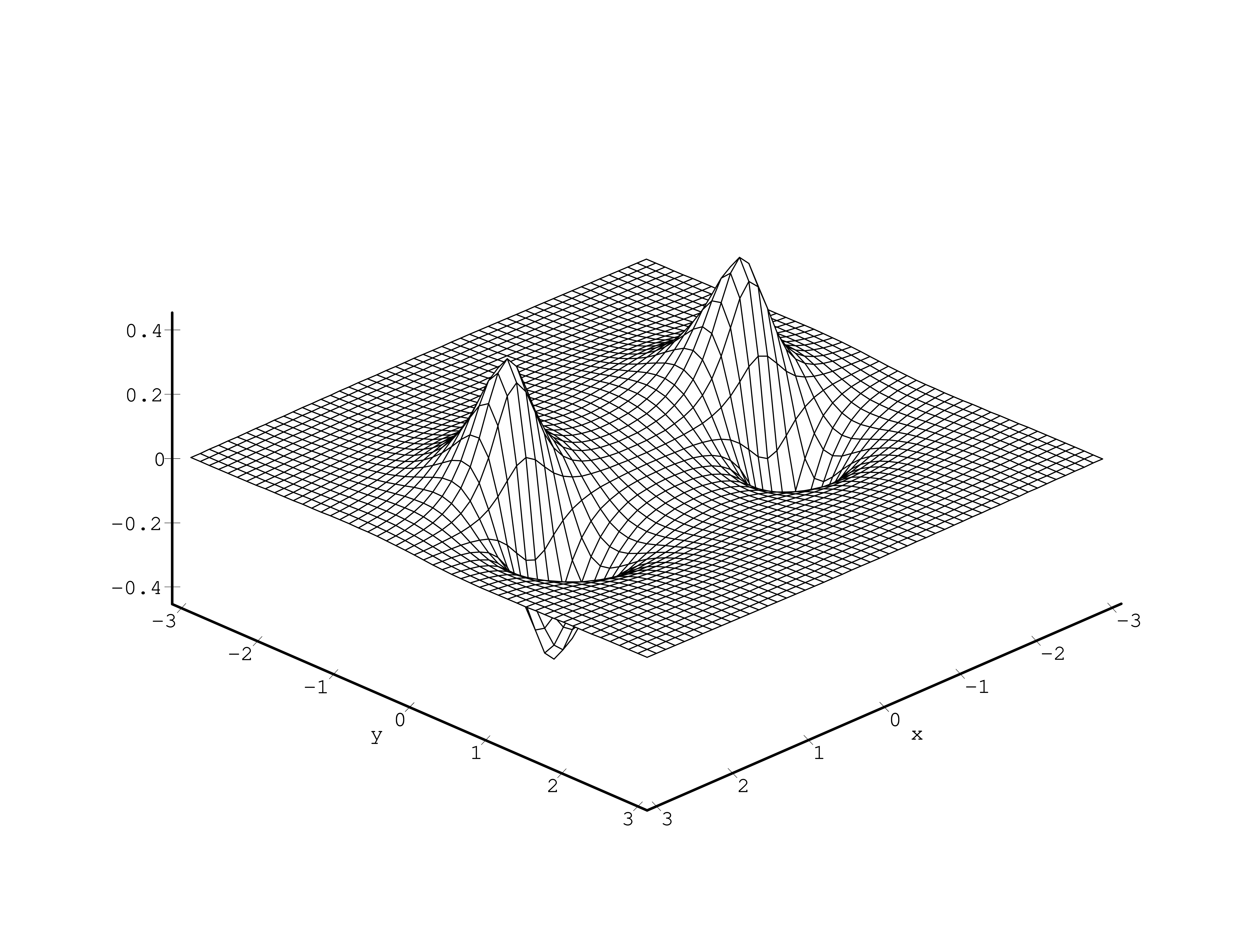}
\end{picture}
\par
\hfill
\begin{picture}(9,3)
\put(2,2.75){$t=1$}
\epsfxsize=10cm
\epsffile{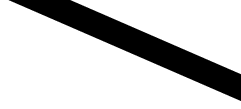}
\end{picture}
\par
\hfill
\put(,5.5){(b)}
\begin{picture}(16,)
\put(2,2.75){$t=-2$}
\epsfxsize=10cm
\epsffile{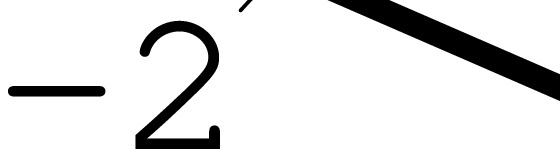}
\end{picture}
\par
\hfill
\begin{picture}(9,3)
\put(2,2.75){$t=2$}
\epsfxsize=10cm
\epsffile{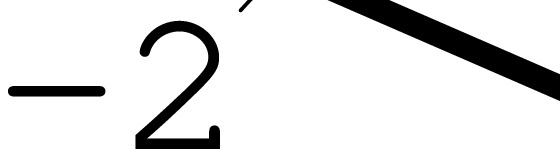}
\end{picture}
\par  
\hfill
\put(,5.5){(c)}
\begin{picture}(16,2)
\put(2,2.75){$t=-5$}
\epsfxsize=10cm
\epsffile{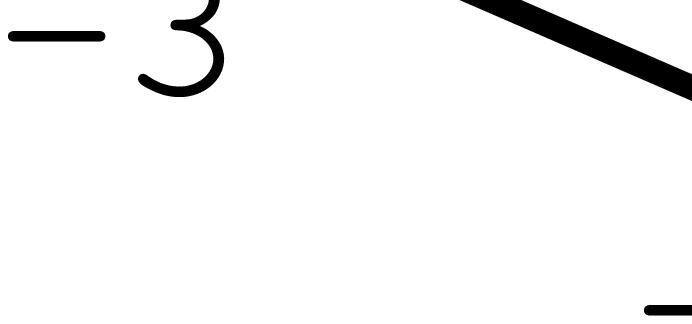}
\end{picture}
\par  
\hfill
\begin{picture}(9,3)
\put(2,2.75){$t=5$}
\epsfxsize=10cm
\epsffile{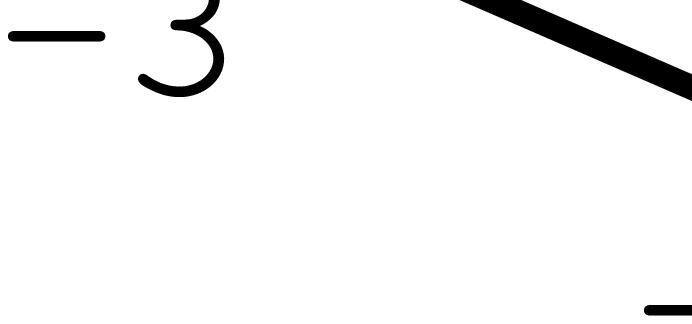}
\end{picture}
\end{figure}

{\bf Acknowledgements.}
TI acknowledges support from EU ERBFMBICT950035.

\hspace{-6mm}{\bf Figure Captions.}\acc
{\bf Figure 1:} The energy density $P^0$ (17) at increasing times 
for (a) $\lm=0$ and (b) $\lm=2$.\acc
{\bf Figure 2:}  The $y$-momentum $P^2$ (18) at various times for (a) 
$\lm=0$ (b) $\lm=2$ and (c) $\lm=10$.

\end{document}